\documentclass[twocolumn,trackchanges]{aastex6}
\RequirePackage{lineno}
\usepackage{amsmath}
\usepackage{amssymb}
\usepackage{amsthm}
\usepackage{natbib,aasdefs,url,bm}
\usepackage{array}
\usepackage{float}
\usepackage{graphicx}
\usepackage{subfigure}
\usepackage{color}

\newcounter{ichi}
\newcounter{ni}
\newcounter{san}
\newcounter{yon}
\def\be{\begin{equation}}
\def\ee{\end{equation}}
\def\ba{\begin{eqnarray}}
\def\ea{\end{eqnarray}}

\newcommand{\ergcmsqs}{\mbox{erg~cm$^{-2}$~s$^{-1}$}}
\newcommand{\gevcmsqs}{\mbox{GeV~cm$^{-2}$~s$^{-1}$}}
\newcommand{\diffuse}{\mbox{GeV~cm$^{-2}$~s$^{-1}$~sr$^{-1}$}}

\def\aap{\ {A\&A}\ }

\def\apjl{\ {ApJL}\ }
\def\apjs{\ {ApJS}\ }

\shorttitle{Blazar Flares as an Origin of High-Energy Cosmic Neutrinos?}
\shortauthors{Murase, Oikonomou, Petropoulou}

\begin{document}

\title{Blazar Flares as an Origin of High-Energy Cosmic Neutrinos?}




\author{Kohta Murase\altaffilmark{1,2,3,4}, Foteini Oikonomou\altaffilmark{5}, Maria Petropoulou\altaffilmark{6}, }
\altaffiltext{1}{Department of Physics, Pennsylvania State University, University Park, PA 16802, USA}
\altaffiltext{2}{Department of Astronomy \& Astrophysics, Pennsylvania State University, University Park, PA 16802, USA}
\altaffiltext{3}{Center for Particle and Gravitational Astrophysics, Pennsylvania State University, University Park, PA 16802, USA}
\altaffiltext{4}{Center for Gravitational Physics, Yukawa Institute for Theoretical Physics, Kyoto University, Kyoto, Kyoto 606-8502, Japan}
\altaffiltext{5}{European Southern Observatory, Karl-Schwarzschild-Str. 2, Garching bei M\"unchen D-85748, Germany}
\altaffiltext{6}{Department of Astrophysical Sciences, Princeton University, Princeton, NJ 08544, USA}


\begin{abstract}
We consider implications of high-energy neutrino emission from blazar flares, including the recent event IceCube-170922A and the 2014-2015 neutrino flare that could originate from TXS 0506+056. 
First, we discuss their contribution to the diffuse neutrino intensity taking into account various observational constraints.
Blazars are likely to be subdominant in the diffuse neutrino intensity at sub-PeV energies, and we show that blazar flares like those of TXS 0506+056 could make $\lesssim1-10$\% of the total neutrino intensity. 
We also argue that the neutrino output of blazars can be dominated by the flares in the standard leptonic scenario for their $\gamma$-ray emission, and energetic flares may still be detected with a rate of $\lesssim1~{\rm yr}^{-1}$.  
Second, we consider multi-messenger constraints on the source modeling. 
We show that luminous neutrino flares should be accompanied by luminous broadband cascade emission, emerging also in X-rays and $\gamma$-rays.
This implies that not only $\gamma$-ray telescopes like {\it Fermi} but also X-ray sky monitors such as {\it Swift} and {\it MAXI} are critical to test the canonical picture based on the single-zone modeling. 
We also suggest a two-zone model that can naturally satisfy the X-ray constraints while explaining the flaring neutrinos via either photomeson or hadronuclear processes.  
\end{abstract}


\section{Introduction}
Recent discoveries of high-energy cosmic neutrinos and gravitational waves have opened up a new era of multi-messenger particle astrophysics~\citep{Aartsen:2013bka,Aartsen:2013jdh,Abbott:2016blz,TheLIGOScientific:2017qsa,Monitor:2017mdv}.
Whereas gravitational wave sources have been detected as individual events, no high-energy neutrino source has been confirmed so far.
The observed diffuse neutrino intensity can be regarded as an isotropic neutrino background (INB) produced by a large number of sources beyond our Galaxy, because the Galactic contribution has been shown to be subdominant \citep[for a review, see][]{Halzen:2016gng}. The origin of cosmic neutrinos is under active debate. 

What is the fastest way to find the neutrino sources individually? Transient sources are the most promising targets, because the atmospheric background can be largely reduced by taking advantage of the time and space coincidences. The brightest transients are detectable with current detectors such as IceCube and KM3Net, even if their contribution to the INB is subdominant. 
Perhaps the most well-known example of neutrino-candidate transients is the prompt emission from $\gamma$-ray bursts (GRBs) with a typical duration of $\sim1-100$~s~\citep[e.g.,][]{Waxman:1997ti,Murase:2005hy,Petropoulou:2014lja,Bustamante:2014oka}, although low-power GRBs have a longer duration of $10^3-10^4$~s~\citep[][]{Murase:2006mm,Gupta:2006jm,Murase:2013ffa}. Others include GRB afterglows~\citep[][]{Waxman:1999ai,Murase:2007yt,Razzaque:2013dsa}, supernovae~\citep[][]{Murase:2010cu,Murase:2017pfe,Petropoulou:2017ymv}, tidal disruption events~\citep[e.g.,][]{Murase:2008zzc,Wang:2011ip}, microquasars~\citep[e.g.,][]{Levinson:2001as,Distefano:2002qw,Torres:2004tm}, and blazar flares~\citep[e.g.,][]{bednarek_1999,Atoyan:2001ey,Halzen:2005pz,Dermer:2012rg,Dermer:2014vaa,Petropoulou:2016ujj,Gao:2016uld}.  

Blazars, a subclass of active galactic nuclei (AGNs) with relativistic jets pointing toward the observer \citep{Urry:1995mg}, and their misaligned counterpart, radio galaxies, have been discussed as the sources of ultrahigh-energy cosmic rays (UHECRs) and/or high-energy neutrinos~\citep[see][for a review]{Murase:2015ndr}. 
Blazars are classified into BL Lac objects (BL Lacs) and quasar-hosted blazars that are mostly flat-spectrum radio quasars (FSRQs). Blazars can also be divided into high-synchrotron-peaked (HSP), intermediate-synchrotron-peaked (ISP), and low-synchrotron-peaked (LSP) objects. 
The acceleration and survival of UHECR nuclei are possible in BL Lacs~\citep{Murase:2011cy,Rodrigues:2017fmu}, whereas efficient photodisintegration and neutrino production are expected in FSRQs~\citep{Murase:2014foa,Palladino:2018lov}. 

Recently, \cite{Aartsen2018blazar1} have reported a $\sim0.1-1$~PeV muon neutrino event, IceCube-170922A, coincident with a month- to year-long $\gamma$-ray flare of the blazar TXS 0506+056 at redshift $z\approx0.336$~\citep{Paiano:2018qeq}. The public alert was sent via the Astrophysical Multi-messenger Network Observatory (AMON), and the follow-up searches led to the discovery of GeV-TeV $\gamma$-ray counterparts as well as X-ray and optical emission~\citep{Aartsen2018blazar1,Keivani:2018rnh}. 
Furthermore, the archival search of the past IceCube data revealed $13\pm5$ signals of lower-energy muon neutrinos coming from the same region in the sky on a time scale of 5 months~\citep{Aartsen2018blazar2}.  
Although it is still too early to be conclusive about their physical association, the reported significance of $\sim3\sigma-4\sigma$ is interesting enough to make us discuss the implications of the neutrino flare-blazar connection.  

{\it IceCube-170922A and the 2014-2015 neutrino flare:}
The neutrino energy estimated for IceCube-170922A is $E_\nu\sim0.3$~PeV, and the $p$-value (chance probability) for the coincidence with the flare from the ISP/LSP blazar, TXS 0506+56, is $\sim0.3$\%, corresponding to a significance of $\approx3\sigma$~\citep{Aartsen2018blazar1}. 
The neutrino flare found in the lower-energy data prior to the discovery of IceCube-170922A has a significance of $\approx3.5~\sigma$~\citep{Aartsen2018blazar2}.
The inferred muon neutrino energy fluences are $E_\nu^2\phi_{\nu_\mu}\sim{10}^{-4}-{10}^{-3}~{\rm erg}~{\rm cm}^{-2}$, implying a released neutrino energy of ${\mathcal E}_\nu^{\rm fl}\sim{10}^{53}-{10}^{54}$~erg. With a flare duration of $t_{\rm dur}\sim10^{7}$~s, the flaring neutrino luminosity is estimated to be $L_{\nu}^{\rm fl}\sim{10}^{46}-{10}^{47}~{\rm erg}~{\rm s}^{-1}$, comparable to the $\gamma$-ray luminosity of the 2017 flare of TXS 0506+56, $L_\gamma^{\rm fl}\sim2\times{10}^{47}~{\rm erg}~{\rm s}^{-1}$ at $0.1-300$~GeV.

{\it IceCube-160731:}
A high-energy track event with an energy higher than several hundred TeV was coincident with the $\gamma$-ray counterpart detected by {\it AGILE}, AGL J1418+0008~\citep{Lucarelli:2017hhh}. The $\gamma$-rays were seen $1-2$ days before IceCube-160731, with a possible association with the BL Lac object, 1RXS J141658.0-001449.

{\it Big Bird (HESE-35):}
This high-energy starting event had a deposited energy of 2~PeV, which could be associated with the FSRQ, PKS B1424-418 at $z=1.522$. Whereas the angular uncertainty for such shower events is $\sim10^\circ-15^\circ$, the $p$-value for the coincidence was 0.05~\citep{Kadler:2016ygj}. If this association is physical, the estimated neutrino luminosity is $L_\nu\gtrsim3\times{10}^{48}~{\rm erg}~{\rm s}^{-1}$. 

This work focuses on implications of IceCube-170922A and the 2014-2015 neutrino flare, assuming that the association with TXS 0506+056 is physical. 
We first examine the connection between blazars and the INB (Sec.~\ref{sec:INB}), and argue that neutrino flares like the ones observed from TXS 0506+056 in 2017 and 2014-2015 are likely to be rare and bright events.
We then show that X-ray observations are critical in testing the standard blazar scenario for neutrino emission and for explaining either flare event observed from this blazar, and we discuss possible multizone models in Section~\ref{sec:models}. 
We conclude in Section~\ref{sec:summary}.

\section{Contribution to the INB}\label{sec:INB}
Here we discuss existing constraints on the blazar contribution to the INB, which are obtained for time-averaged emission. Given that blazars are variable sources across the electromagnetic spectrum, we then investigate the contribution of blazar flares to the INB.

\subsection{General Constraints} 
The blazar contribution to the INB has been constrained by different types of analyses: 
(i) diffuse searches for extremely high-energy (EHE) neutrinos~\citep{Aartsen:2016ngq,Aartsen:2017mau}, 
(ii) event clustering and autocorrelation analyses~\citep{Murase:2016gly,Aartsen:2014ivk,Aartsen:2017kru}, and (iii) stacking and cross-correlation analyses~\citep{Aartsen:2016lir,Aartsen:2017kru}. 

Neutrino spectra have been predicted by most theoretical models to be hard and peaking at energies beyond 1 PeV. The hardness of the spectrum is related to the fact that the target photon density in blazars is higher at lower energies. Upon normalization to the IceCube flux at $\sim 1$ PeV, most model-predicted fluxes at 10 PeV are found to be $E_\nu^2\Phi_\nu\gtrsim(3-5)\times{10}^{-8}~\diffuse$ for all flavors~\citep[][and references therein]{Murase:2015ndr}. With the 9~yr diffuse analysis, the IceCube Collaboration reported an upper limit on the INB, $E_\nu^2\Phi_\nu\lesssim1\times{10}^{-8}~\diffuse$ (90\% CL) at $5-10$~PeV~\citep{Aartsen:2017mau}, excluding some of the optimistic physical models for blazar neutrino emission~\citep[see also][]{Neronov:2016ksj}. 

Another type of less model-dependent constraints is obtained from the absence of sources of high-energy multiplets ~\citep{Murase:2016gly,Ahlers:2014ioa} \citep[see also][for earlier works]{Lipari:2008zf,Silvestri:2009xb,Murase:2012df}. Let the number of the sources with multiplets be denoted as $N_{m\geq k}$. Then, constraints can be placed by requiring $N_{m\geq k}\leq1$ at sufficiently high energies (e.g., $>50$~TeV for muons). Although the energy-dependent effective area should be taken into account for detailed calculations as in \cite{Murase:2016gly}, the basic results can be understood by using a limit from the nondetection of point sources. The 8~yr point-source sensitivity (90\% CL) for an $E^{-2}$ neutrino spectrum is $F_{\rm lim}\sim(5-6)\times{10}^{-10}$~\gevcmsqs~\citep{Aartsen:2017kru}. 
For such a flat energy spectrum with a time-averaged luminosity of $\varepsilon_\nu L_{\varepsilon_{\nu_\mu}}^{\rm ave}\sim10^{44}~{\rm erg}~{\rm s}^{-1}$, the number density of the sources is constrained as:
\begin{eqnarray}\label{eq:n0a}
n_0^{\rm eff}&\lesssim&1.9\times10^{-10}~{\rm Mpc^{-3}}~\left(\frac{\varepsilon_\nu L_{\varepsilon_{\nu_\mu}}^{\rm ave}}{{10}^{44}~{\rm erg\,s^{-1}}}\right)^{-3/2}\nonumber\\
&\times&{(b_mq_L/6.6)}^{-1}F_{\rm lim,-9.2}^{3/2}\left(\frac{2\pi}{\Delta\Omega}\right),
\end{eqnarray}
where $q_L\sim1-3$ is a luminosity-dependent correction factor determined by the redshift evolution and $\Delta \Omega$ is the solid angle covered by the detector. 
This limit depends on spectral templates of the sources, and the differential sensitivity can be worse than the integrated sensitivity by $\sim\ln(10)$. Thus, for harder spectra explaining neutrinos only in the PeV range~\citep[see Figs.~2 and 4 of][]{Murase:2016gly}, the upper limit in Eq.~(\ref{eq:n0a}) can be relaxed by $\sim3$.
Note that $b_m\geq1$ is a factor that represents details of the analysis. For example, in the case of the multiplet analysis, we have $b_m\approx6.6$ for $m\geq2$ and $b_m\approx1.6$ for $m\geq3$. 

Using Eq.~(\ref{eq:n0a}) we derive an upper limit on the contribution of a source population with $n_0^{\rm eff}$ and to the INB:
\begin{eqnarray}\label{eq:diffuse}
E_\nu^2\Phi_\nu&\approx&\frac{\xi_z ct_H}{4\pi}3(\varepsilon_\nu L_{\varepsilon_{\nu_\mu}}^{\rm ave})n_0^{\rm eff}\nonumber\\
&\lesssim&6.9\times{10}^{-9}~\diffuse~\left(\frac{\xi_z}{0.7}\right){\left(\frac{6.6}{b_mq_L}\right)}^{2/3}\nonumber\\
&\times&\left(\frac{n_0^{\rm eff}}{{10}^{-7}~{\rm Mpc}^{-3}}\right)^{1/3}F_{\rm lim,-9.2}{\left(\frac{2\pi}{\Delta\Omega}\right)}^{2/3},
\end{eqnarray}
where $t_H$ is the Hubble time and $\xi_z$ represents the redshift evolution of the neutrino luminosity density of the sources: $\xi_z\sim0.7$ for the $\gamma$-ray luminosity density evolution of BL Lacs, $\xi_z\sim8$ for that of FSRQs, and $\xi_z\sim3$ for the X-ray luminosity density evolution of AGNs~\citep{Ajello:2013lka,Ueda:2014tma}. Note that if one uses the number density evolution (corresponding to the equal luminosity weight), we have $\xi_z\sim0.2$ for all BL Lacs with $n_s\propto{(1+z)}^{-3.5}$ and $\xi_z\sim0.1$ for HSP objects with $n_s\propto{(1+z)}^{-6}$, respectively~\citep{Ajello:2013lka}. 
In the extreme case where the faintest BL Lacs are equally neutrino emitters, we have $n_0^{\rm eff}=n_0^{\rm tot}\sim(1-3)\times{10}^{-7}~{\rm Mpc}^{-3}$~\citep{Ajello:2013lka} and $\xi_z\sim0.2$, implying that the contribution to the INB is $\lesssim10$\% at 0.1~PeV. Although the limit could be further relaxed by a factor of two by integrating the number density down to the faintest source tail, blazars are unlikely to be dominant in the INB for such weak redshift evolution. One should keep in mind that the effective number density, which depends on the neutrino luminosity function (i.e., $dn_s/dL_\nu$), should always be smaller than the total number density, i.e., $n_0^{\rm eff}<n_0^{\rm tot}$~\citep{Murase:2016gly}. 
This is because physical models typically predict $L_{\nu}\propto L_\gamma-L_\gamma^2$~\citep{Murase:2014foa,Tavecchio:2014eia,Petropoulou:2015upa,Murase:2016gly}.
In the fiducial case of the leptonic scenario for BL Lacs, the effective number density is $n_0^{\rm eff}\sim{10}^{-9}-{10}^{-8}~{\rm Mpc}^{-3}$, which gives $\lesssim(5-10)(\xi_z/0.7)$\% of the INB. 
For FSRQs, we have $n_0^{\rm eff}\sim{10}^{-12}-{10}^{-11}~{\rm Mpc}^{-3}$, leading to $\lesssim(6-10)(\xi_z/8)$\% in the 0.1~PeV range. 
Note that the neutrino luminosity density evolution would be stronger than the $\gamma$-ray one if the $\gamma$-ray luminosity more strongly weighs on the neutrino luminosity, but such cases are constrained by the stacking limits. In this sense these constraints are complementary to each other. 

Another limit can be placed by the autocorrelation analysis on the small-scale anisotropy~\citep{Aartsen:2014ivk,Aartsen:2017kru,Ando:2017xcb}.  With the measured INB and the latest anisotropy limit~\citep{Aartsen:2017kru}, the upper limit on the Poisson angular spectrum is estimated to be $E_\nu^4C_P<4\times{10}^{-19}~{\rm GeV}^2~{\rm cm}^{-2}~{\rm s}^{-2}~{\rm sr}^{-1}$. Then, using the known formula of $C_P$ for standard candle sources, we obtain $n_0^{\rm eff}\lesssim1.1\times10^{-9}~{\rm Mpc^{-3}}~{(\varepsilon_\nu L_{\varepsilon_{\nu_\mu}}^{\rm eff}/{10}^{44}{\rm erg\,s^{-1}})}^{-3/2}q_L^{-1}{(2\pi/\Delta\Omega)}^{1/4}$, which gives a comparable limit. 
 
Blazars are highly variable objects, and as a result, their luminosity density could be dominated by the flaring states characterized by a ``flaring'' neutrino luminosity $L_{\nu}^{\rm fl}$. In this ``flare-dominated'' case, the average neutrino luminosity can be written as $L_\nu^{\rm ave}\approx f_{\rm fl}L_\nu^{\rm fl}$, where $f_{\rm fl}$ is the duty factor of flares, which will be discussed in the next subsection. The differential neutrino luminosity density is then written as $\varepsilon_\nu Q_{\varepsilon_\nu}=(\varepsilon_\nu L_{\varepsilon_\nu}^{\rm ave})n_0^{\rm eff}\approx(\varepsilon_\nu L_{\varepsilon_\nu}^{\rm fl})(f_{\rm fl}n_0^{\rm eff})$. For transients, including flaring sources, the atmospheric background can be reduced owing to the shorter time window, thereby improving, in general, the fluence sensitivity. The power of such a time-dependent search was demonstrated in \cite{Aartsen2018blazar2} (although the excess besides IceCube-170922A was not significant in the time-integrated search).
Thus, Eq.~(\ref{eq:n0a}) for the time-averaged emission can still be regarded as a ``conservative'' limit on flaring sources, and the constraint given by Eq.~(\ref{eq:diffuse}) is still applicable even if blazars are highly variable\footnote{Note that multiplet limits are applicable to transients, and we stress that such a ``transient'' case is here considered. Indeed, even if the 2014 flare of TXS 0506+056 is included, $N_{m\geq k}$ for $\gtrsim50$~TeV muons seems no more than one, leading to, e.g., $\rho_{0}^{\rm eff}\gtrsim1.7\times10^{4}\,{\rm Gpc^{-3}}\,{\rm yr^{-1}}\,{(b_mq_L/6.6)}^2{({\Delta \Omega/2\pi)}^2{(T_{\rm obs}/8~{\rm yr})}^2}{(\xi_z/0.7)}^{-3}\phi^{-3}_{\rm lim, -1}$\\${\rm max}[N_{\rm fl},1]$ for transients accounting for all the diffuse flux. Here $N_{\rm fl}\approx f_{\rm fl}T_{\rm obs}/t_{\rm dur}\approx T_{\rm obs}/\Delta T_{\rm fl}$ is the number of flaring periods and $\Delta T_{\rm fl}$ is the typical flare interval. When at least one flare occurs in the observation time (i.e., $\Delta T_{\rm fl}\lesssim T_{\rm obs}$), the expressions for density and diffuse limits on flaring neutrino sources eventually become similar to those for steady sources, and one can easily obtain Eqs.~(\ref{eq:n0a}) and (\ref{eq:diffuse}). Here substituting the time-averaged sensitivity gives conservative results because of $F_{\rm lim}>\phi_{\rm lim}/T_{\rm obs}$, but this does not mean that we assume steady sources. The constraints for rare transients such that $\Delta T_{\rm fl}\gtrsim T_{\rm obs}$, including non-repeating ones, can be obtained by replacing $n_{0}^{\rm eff}$ in Eq.~(\ref{eq:diffuse}) with $\rho_0^{\rm eff}T_{\rm obs}\approx n_{0}^{\rm eff}(T_{\rm obs}/\Delta T_{\rm fl})$. Because of $n_{0}^{\rm eff}(T_{\rm obs}/\Delta T_{\rm fl})\lesssim n_{0}^{\rm eff}$, Eq.~(\ref{eq:diffuse}) still gives a conservative limit on the diffuse flux contribution.} 
. 

As noted above, stacking and cross-correlation analyses can provide tighter constraints especially for physically motivated models. In particular, for the 2LAC catalog consisting of 862 blazars, the blazar contribution to the INB is restricted to be $\leq(19-27)$\%~\citep{Aartsen:2016lir}. Note that theory typically predicts $L_\nu\propto L_\gamma-L_\gamma^2$~\citep{Murase:2014foa,Tavecchio:2014eia,Petropoulou:2015upa}, and a flux weighting with $F_\nu\propto F_\gamma$ leads to $\lesssim7$\%~\citep{Aartsen:2016lir}.  For HSP BL Lacs, the preliminary result gives $\leq(4.5-5.7)$\%~\citep{Aartsen:2017kru} in the equal flux weight ($F_\nu\propto {\it const.}$) assumption. The stacking limits are powerful and more meaningful when the physical luminosity weight is strong \cite[in which the neutrinos mostly come from the resolved blazars; see][]{Murase:2014foa}. Thus, $\gamma$-ray-bright blazars that significantly contribute to the extragalactic $\gamma$-ray background are disfavored as the dominant ($\sim100$\%) origin of IceCube's neutrinos \citep[see also][]{Wang:2015woa,Zhang:2016vbb,Neronov:2016ksj,Palladino:2017aew,Ando:2017xcb}. 

All the constraints discussed so far can be relaxed by a factor of a few under different assumptions.   
(i) The diffuse EHE limits can be avoided if the cosmic-ray (CR) spectrum is sufficiently soft or the maximum CR energy is lower than $\sim10-100$~PeV (far below UHECR energies), as considered in \cite{Murase:2014foa} and \cite{Dermer:2014vaa}. This is because the neutrino production efficiency increases with energy, and the resulting neutrino spectra are hard for a power-law CR spectrum with $s_{\rm cr}\sim2-2.6$. 
(ii) The multiplet limit is sensitive to $\xi_z$. Weakly evolving sources such as BL Lacs are strongly constrained. On the other hand, rapidly evolving sources such as FSRQs could give a significant contribution to the INB if $n_0^{\rm eff}\sim n_0^{\rm tot}\sim{10}^{-9}~{\rm Mpc}^{-3}$. However, this is contrary to the fiducial theoretical prediction, $n_0^{\rm eff}\ll n_0^{\rm tot}$. \cite{Dermer:2014vaa} proposed such a model, in which flaring blazars significantly contribute to the INB only in the PeV range, but this model does not explain the UHECRs~\citep[see also][]{Murase:2015ndr}. 
(iii) The stacking limits do not apply to $\gamma$-ray dark blazars. For example, a subset of FSRQs with a spectral energy distribution (SED) peak in the MeV range are dim in the {\it Fermi} LAT band, and such MeV blazars could significantly contribute to the INB as {\it hidden} CR accelerators~\citep{Murase:2015xka}. 

The combination of all constraints indicates that the blazar contribution to the INB is likely to be subdominant at least in the 0.1~PeV range. This is even more so the case for the medium-energy component in the $10-100$~TeV range, which requires models prohibiting the escape of $\gamma$-rays~\citep{Murase:2015xka}. On the other hand, at present, the contribution could be more significant in the PeV or higher-energy range. We note that fiducial models (normalized to the UHECR flux), presented in \cite{Murase:2014foa}, are consistent with the above constraints, and give $\sim2-10$\% of the INB in the 0.1~PeV range (and more at higher energies) .  

\subsection{Implications of TXS 0506+056}
Bright flaring sources are detectable in neutrinos whether the blazars are dominant or subdominant in the extragalactic neutrino sky~\citep[][]{Dermer:2014vaa,Murase:2016gly,guepin_2017}. 
Nevertheless, it is natural to estimate the contribution of blazar flares like the ones from TXS 0506+056 to the INB and discuss the detectability of similar flaring events.

The flaring state lasts only for a fraction of the observation time. For a given time binning, one can measure the number of detected particles (e.g., photons) per bin, which is proportional to luminosity. 
We introduce the flaring state when the number of photons in a bin exceeds a certain threshold corresponding to the luminosity $L^{\rm th}$. Then one can construct the distribution of the number of time bins with luminosity, $dN/dL$.
The fraction of time spent in the flaring state (i.e., the duty factor) is given by
\begin{equation}
\label{eq:fl}
f_{\rm fl}=\frac{1}{N_{\rm tot}}\int_{L^{\rm th}}dL\, \frac{dN}{dL},
\end{equation}
where $N_{\rm tot}$ is the total number of time bins. 
The fraction of energy emitted in the flaring state is:
\begin{equation}
\label{eq:bfl}
b_{\rm fl}=\frac{1}{L^{\rm ave}N_{\rm tot}}\int_{L^{\rm th}}dL\, L\frac{dN}{dL},
\end{equation}
where the average luminosity is given by $L^{\rm ave}=(1/N_{\rm tot})\int dL\, L(dN/dL)$ and the average flaring luminosity is rewritten as $L^{\rm fl}=(b_{\rm fl}/f_{\rm fl})L^{\rm ave}$. Note that in the flare-dominated limit, $b_{\rm fl}\approx1$, we have $L^{\rm ave}\approx f_{\rm fl}L^{\rm fl}$.    

\begin{figure}[t]
\includegraphics[width=\linewidth]{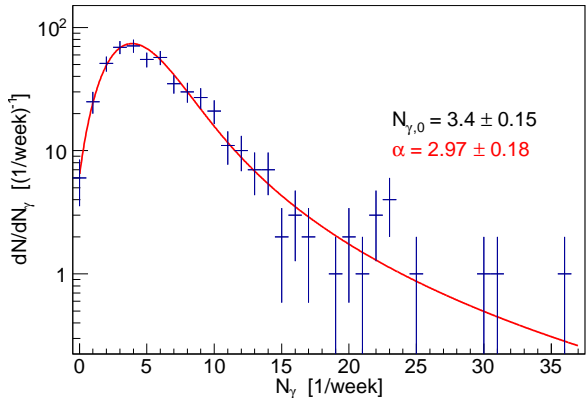}
\caption{Histogram of the number of photons, $N_{\gamma}$ detected per week by the FAVA analysis in the direction of TXS 0506+056 in the high-energy bin ($800$~MeV$-$$300$~GeV). 
The photon distribution is modeled as a power law with spectral index $\alpha$ above a minimum ``quiescent" count rate $N_{\gamma,0}$ (red solid line). The error bars are statistical. 
\label{fig1}
}
\vspace{-1.\baselineskip}
\end{figure}

Using the public {\it Fermi} All-sky Variability Analysis (FAVA) data by {\it Fermi} LAT~\citep{2017ApJ...846...34A}, one can obtain the luminosity distribution, $dN/dL_\gamma$ of a source, under the assumption that the spectral shape does not change during flaring states (since only photon counts are available in the FAVA analysis). In the latter case, $dN/dL_\gamma \propto dN/dN_\gamma$. Henceforth, we use $dN/dN_{\gamma}$ and $dN/dL_{\gamma}$ interchangeably. 
An example for TXS 0506+056 is shown in Fig.~\ref{fig1}, in which the photon distribution is modeled as a power law with slope $\alpha$ above a minimum number of photons per bin $N_{\gamma,0}$ that corresponds to a ``quiescent'' flux. 
The number of detected photons per time interval is then given by a convolution of this power law with a Poissonian distribution. We applied the same method to a selection of BL Lacs at intermediate redshifts from FAVA, and find that they are well described by a power-law $dN/dL_\gamma\propto L_\gamma^{-\alpha}$ with $\alpha\sim2-4$.  
One can also calculate $f_{\rm fl}$ and $b_{\rm fl}$ for various blazars using Eqs.~(\ref{eq:fl}) and (\ref{eq:bfl}) (see Table~\ref{tb1}). The exact values depend on the definition of the flaring state (see, e.g., \citealp{2009A&A...502..499R}), but our main conclusions are not expected to change, if flares are defined differently. 
We found that the duty factor lies in the range of $0.3-10$\% for $\geq5\sigma$ flares (as per the FAVA definition), and obtained $f_{\rm fl}\approx0.02-0.1$ 
for TXS 0506+056. The corresponding fraction of emitted photons is $b_{\rm fl}\sim0.1\ll1$, implying that the $\gamma$-ray emission is dominated by steady emission. Although a detailed statistical study is beyond the purpose of the present work, we underline the need for a systematic study of the properties of flaring, $\gamma$-ray-bright blazars. For the purposes of the present work, we treat TXS 0506+056 as a ``characteristic'' test case.

During the flare of TXS 0506+056  in the period 2017 September 15-27, the $1-100$ GeV ``differential'' $\gamma$-ray luminosity ($\varepsilon_\gamma L_{\varepsilon_\gamma}^{\rm fl}\sim2\times{10}^{46}~{\rm erg}~{\rm s}^{-1}$) was about 6 times higher than the average in the 3FGL catalog ($\varepsilon_\gamma L_{\varepsilon_\gamma}^{\rm ave}\sim4\times{10}^{45}~{\rm erg}~{\rm s}^{-1}$) \citep{2017ATel10791....1T,Acero:2015hja}.\footnote{In the {\it Fermi} ATel, a different energy range is used for the {\it Fermi} analysis ($0.1-300$~GeV), giving $L_{\gamma,0.1 - 300~{\rm GeV}}^{\rm fl}\sim2\times{10}^{47}~{\rm erg}~{\rm s}^{-1}$ for the $\gamma$-ray  luminosity of TXS 0506+056 during the flare. Correspondingly for the quiescent luminosity in the 3FGL catalogue we obtain $L_{\gamma,0.1 - 300~{\rm GeV}}^{\rm ave}\sim3\times{10}^{46}~{\rm erg}~{\rm s}^{-1}$ from the flux estimate of \citet{2017ATel10791....1T} assuming an unbroken power-law with index $\beta = 2.0$.} 
Thus this flare can be regarded as one of the brightest flares of this object. 
Note that the $\gamma$-ray photon index in the $0.1-300$~GeV range had a similar value during the September flare, $\beta=2.0$, to that in the 3FGL. 

\begin{table}[t]
\caption{Flare Duty Factors ($f_{\rm fl}$) and Flare Energy Fractions ($b_{\rm fl}$) for TXS 0506+056, OJ 287, PKS 0426-380, PKS 0301-243, S5 0716+071 and S4 0954+065 as Derived from the FAVA Analysis~\citep{2017ApJ...846...34A}.
}
\label{tb1}
\begin{center}
\small
\begin{tabular}{|c||c|c|c|c|c|c|}
\hline 
Name & $L_{\gamma}$ & $f_{\rm fl}^{\rm LE}$ & $f_{\rm fl}^{\rm HE}$ & $b_{\rm fl}^{\rm LE}$ & $b_{\rm fl}^{\rm HE}$ & $\alpha$\\
\hline 
TXS 0506+056 & ${10}^{46.3}$ & 0.1 & 0.03 & 0.1 & 0.1 & $3.0$\\
\hline
OJ 287 & ${10}^{46.1}$ & 0.02 & 0.01 & 0.04 & 0.1 & $2.9$ \\
\hline
PKS 0426-380 & ${10}^{48}$ & $0.1$ & $0.1$ & 0.2 & 0.2 & $1.7$\\
\hline
PKS 0301-243 & ${10}^{46}$ & $0.04$ & $0.05$ & 0.1 & 0.3 & $2.5$\\
\hline
S5 0716+071 & ${10}^{46.7}$ & 0.07 & 0.08 & 0.1 & 0.2 & $1.7$\\
\hline
S4 0954+065 & $10^{45.5}$ & 0.04 & 0.03 & 0.07 & 0.3 & $2.5$\\
\hline
\end{tabular}
\end{center}
{\bf Note.} Values are reported for the low-energy (LE: $100-800$~MeV) and high-energy (HE: $800$~MeV$-$$300$~GeV) bins. The duty factors quoted are for flaring periods when the flux is enhanced by $\geq 5\sigma$ according to the FAVA definition. The ``integrated'' $\gamma$-ray luminosity of the sources, $L_{\gamma}$ (in units of ${\rm erg}~{\rm s}^{-1}$), is derived from the 3FGL in the $1-100$~GeV energy range.
Rounded values of the power-law index ($\alpha$) are also shown. 
\vspace{0.5\baselineskip}
\end{table}

Naively, high-energy neutrino emission is dominated by the non-flaring contribution, if one assumes $L_\nu\propto L_\gamma$. 
However, in the standard leptonic scenario for the blazar $\gamma$-ray emission, flares can dominate the neutrino output of a blazar. Under different assumptions, the leptonic models predict $L_\nu\propto L_\gamma^{\gamma}$ with $\gamma\sim1.5-2$ \citep[see][and references therein]{Murase:2016gly}, giving
\begin{equation}
L_\nu^2 \frac{dN}{dL_\nu}\propto L_\nu^{1-\frac{\alpha-1}{\gamma}},
\end{equation}
which implies that the flaring contribution can be dominant, e.g., for BL Lacs with $\gamma\sim2$ and $\alpha\leq3$. 
Such a situation might be the case for the multi-messenger flare associated with IceCube-170922A. 
In addition, the neutrino flare emission can be pronounced if the CR spectrum is harder during the high state. To demonstrate this, let us consider a toy model where the low-state and high-state CR spectra are described as $\varepsilon_pL_{\varepsilon_p}^{l} \propto\varepsilon_p^{2-s_{l}}$ (for $s_l>2$) and $\varepsilon_pL_{\varepsilon_p}^{\rm fl} \propto \varepsilon_p^{2-s_{\rm fl}}$ (for $s_{\rm fl}<2$), respectively. Assuming that the maximum and minimum energies are $\varepsilon_p^{\rm max} $ and $\varepsilon_p^{\rm min}$, the flux enhancement at $\varepsilon_\nu$ is:
\begin{equation}
c[\varepsilon_\nu]\equiv\frac{\varepsilon_\nu L_{\varepsilon_\nu}^{\rm fl}}{\varepsilon_\nu L_{\varepsilon_\nu}^l}\approx\frac{(2-s_{\rm fl})f_{p\gamma}^{\rm fl}}{(s_l-2)f_{p\gamma}^l}{\left(\frac{20\varepsilon_\nu}{\varepsilon_{p,\rm max}^{\rm fl}}\right)}^{2-s_{\rm fl}}{\left(\frac{20\varepsilon_\nu}{\varepsilon_{p,\rm min}^l}\right)}^{s_l-2},
\end{equation}
where $f_{p\gamma}$ is the effective photomeson production optical depth that we will discuss below. 
If we adopt indicative values, namely, $s_l\sim2.3$, $s_{\rm fl}\sim1.8$, $\varepsilon_\nu\sim0.05\varepsilon_{p,\rm max}^{\rm fl}\sim0.1$~PeV, and $\varepsilon_{p,\rm min}^{l}\sim10$~GeV we find $c[\varepsilon_\nu]\sim30f_{p\gamma}^{\rm fl}/f_{p\gamma}^{l}$. 
Unless the physical conditions during flares change so radically that $f_{p\gamma}^{\rm fl}/f_{p\gamma}^{l}\ll 1$, flares may dominate the neutrino output of a blazar. 
Such a situation could be realized in the 2014-2015 neutrino flare. 
For either case, in the flare-dominated regime, the time-averaged neutrino luminosity can be written as $\varepsilon_\nu L_{\varepsilon_\nu}^{\rm ave}\approx f_{\rm fl}(\varepsilon_\nu L_{\varepsilon_\nu}^{\rm fl})$.


From the above considerations, neutrinos can be copiously produced during the high states. 
In particular, the contribution of blazar flares like those observed from TXS 0506+056 to the INB is constrained as:  
\begin{eqnarray}
E_\nu^2\Phi_\nu&\lesssim&3.8\times{10}^{-10}~\diffuse~\left(\frac{2\pi}{\Delta\Omega}\right)\left(\frac{\xi_z}{0.7}\right)\nonumber\\
&\times&{\left(\frac{0.05}{f_{\rm fl}}\right)}^{1/2}\left(\frac{{10}^{46}~{\rm erg\,s^{-1}}}{\varepsilon_\nu L_{\varepsilon_{\nu_\mu}}^{\rm fl}}\right)^{1/2}\left(\frac{6.6}{b_mq_L}\right)F_{\rm lim,-9.2}^{3/2},\,\,\,\,\,\,\,
\label{eq:diffuse_fl}
\end{eqnarray}
where Eq.~(\ref{eq:diffuse}) is used.
They can contribute up to a few percent of the INB, allowing uncertainties in the redshift evolution and neutrino spectrum.
Eq.~(\ref{eq:diffuse_fl}) also implies that dimmer blazar flares could make a larger contribution, but the total neutrino intensity must not exceed the upper limit given by Eq.~(\ref{eq:diffuse}).

Finally, we estimate how many bright flares can be detected with multi-messenger (neutrino and $\gamma$-ray) coincidence searches in near future. 
For variable sources, such as TXS 0506+056, we can use the muon neutrino fluence sensitivity, $\phi_{\rm lim}\sim0.04~{\rm GeV}~{\rm cm}^{-2}$ (for a spectrum that is broad around 0.1~PeV), which can also be calculated from the public effective area. 
Due to the $\gamma$-ray monitoring of blazars with \textit{Fermi} LAT in GeV energies and HAWC in TeV energies, 
the detection rate of flaring blazars in neutrinos can be estimated to be~\citep[see Eq.~4.14 of][]{Murase:2012df}
\begin{eqnarray}
\dot{N}_{\rm blazar}&\approx&\frac{\Delta\Omega}{3}\rho_0^{\rm eff}{\left(\frac{\varepsilon_\nu L_{\varepsilon_{\nu_\mu}}^{\rm fl}t_{\rm dur}}{4\pi\phi_{\rm lim}}\right)}^{3/2}\nonumber\\
&\lesssim&0.9~{\rm yr}^{-1}~{\left(\frac{f_{\rm fl}}{0.05}\right)}^{-1/2} t_{\rm dur,7}^{1/2} {\left(\frac{F_{\rm lim,-9.2}}{\phi_{\rm lim,-1.4}}\right)}^{3/2}.\,\,\,\,\,
\end{eqnarray}
Here $\rho_0^{\rm eff}=f_{\rm fl}n_{0}^{\rm eff}/t_{\rm dur}$ is the effective rate density of blazar flares, and the above equation is valid when the IceCube observation time is longer than $t_{\rm dur}/f_{\rm fl}$. 
This estimate is consistent not only with existing observational constraints but also with the fact that no other flares besides that of TXS 0506+056 have been identified with a high significance. The prospects of detecting neutrinos from short-duration blazar flares are less favorable, because of $\dot{N}_{\rm blazar}\propto t_{\rm dur}^{1/2}$ \citep[see also][]{Petropoulou:2016ujj,guepin_2017}. If the association with TXS 0506+056 is physical, according to the standard leptonic scenario including FSRQs~\citep{Murase:2014foa,Dermer:2014vaa}, we predict that neutrinos associated with FSRQ flares should also be identified in the future. 
It also suggests that dedicated time-dependent neutrino searches \citep[][]{Turley:2016ver,Turley:2018biv} are important to test these predictions.

\section{Implications for Source Models}\label{sec:models}
\subsection{Importance of X-Ray and $\gamma$-Ray Observations}
\cite{Keivani:2018rnh} provided a detailed study of the TXS 0506+056 flare, using the multi-messenger data that have been obtained quasi-simultaneously with IceCube-170922A. The authors found a viable model for both high-energy neutrinos and $\gamma$-rays only in the leptonic scenario, where $\gamma$-ray emission is attributed to the inverse Compton (IC) mechanism. 
The X-ray and $\gamma$-ray light curves were variable on a day timescale, thus implying a comoving size of the emission region of $l'\approx \delta c t_{\rm var}/(1+z)\simeq4.5\times{10}^{16}~{\rm cm}~(\delta/20)t_{\rm var,5}$, where typical values of the Doppler factor are $\delta\sim10-30$.
The observed SED suggests that the Compton dominance parameter is around unity, suggesting a magnetic field of $B'\sim0.1-1$~G~\citep{Keivani:2018rnh}, which is consistent with population-based estimates for BL Lacs~\citep{celotti_2008, Murase:2014foa}.  

Neutrinos and hadronic $\gamma$-rays are coproduced by photomeson production, which is characterized by its effective optical depth, $f_{p\gamma}$. 
Let us consider a relativistically moving blob and a target photon spectrum, $n_{\varepsilon'_t}$ (where $\varepsilon'_t\approx\varepsilon_t/\delta$ is the target photon energy in the comoving frame). Approximating the spectrum by ${\varepsilon'}_tn_{\varepsilon'_t}=n'_0{({\varepsilon'}_t/{\varepsilon'}_0)}^{1-\beta}$ with $\beta>1$, where ${\varepsilon'}_0$ is the reference energy, $f_{p\gamma}$ is given by~\citep[e.g.,][]{Murase:2015xka}
\begin{equation}\label{eq:fpgamma}
f_{p \gamma}[\varepsilon_p]\approx\eta_{p\gamma}[\beta]\hat{\sigma}_{p\gamma}l'n'_0
{(\varepsilon'_p/\tilde{\varepsilon}_{p\gamma0}^{\prime})}^{\beta-1},
\end{equation}
where $\eta_{p\gamma}[\beta]\approx2/(1+\beta)$, $\hat{\sigma}_{p\gamma}\sim0.7\times{10}^{-28}~{\rm cm}^2$ is the attenuation cross section, $\bar{\varepsilon}_\Delta\sim0.3$~GeV, and $\tilde{\varepsilon}_{p\gamma0}^\prime=0.5m_pc^2\bar{\varepsilon}_{\Delta}/\varepsilon'_0$. 
This estimate is valid when the meson production is dominated by the $\Delta$-resonance and direct pion production. 
For target photons with observed energy $E_t=\varepsilon_t/(1+z)$, the characteristic energy of protons producing neutrinos of energy $\varepsilon_\nu$ is $\varepsilon_p\approx 20\varepsilon_\nu\approx0.5\delta^2m_pc^2\bar{\varepsilon}_\Delta{\varepsilon_t}^{-1}$. This results in $\varepsilon_t\sim8~{\rm keV}~{(\delta/20)}^2{(\varepsilon_\nu/{\rm 300~TeV})}^{-1}$, corresponding to UV photons or X-rays for neutrino energies ranging from $\sim 3$~PeV to $\sim30$~TeV. 
For example, the high-energy synchrotron component of the SED during the TXS 0506+056 flare is well described by $\beta=2.8$ in the optical-to-X-ray range (above the lower-energy peak at $\varepsilon_{\rm syn}\sim1$~eV)~\citep{Keivani:2018rnh}. In this case, for a CR spectrum with $s_{\rm cr}=2$, the predicted neutrino spectrum is so hard (see also Eq.~\ref{eq:fpgamma}) that it would contradict the nondetection of $>10$~PeV neutrinos during the flare, unless the CR proton spectrum cuts off at $10-100$~PeV~\citep{Keivani:2018rnh}.

The same target photons lead to the Bethe-Heitler pair production, to which the effective optical depth is:
\begin{eqnarray}
f_{\rm BH}[\varepsilon_p]&\approx&\eta_{p\gamma}[\beta]\hat{\sigma}_{\rm BH}l'n'_0
{(\varepsilon'_p/\tilde{\varepsilon}_{\rm BH0}^{\prime})}^{\beta-1}\nonumber\\
&=&g[\beta]f_{p\gamma}[\varepsilon_p],
\end{eqnarray}
where $\hat{\sigma}_{\rm BH}\sim0.8\times{10}^{-30}~{\rm cm}^2$, $\bar{\varepsilon}_{\rm BH}\sim10(2m_ec^2)\sim10$~MeV~\citep{1992ApJ...400..181C}, $\tilde{\varepsilon}_{\rm BH0}^\prime=0.5m_pc^2\bar{\varepsilon}_{\rm BH}/\varepsilon'_0$, and $g[\beta]\sim0.011{(30)}^{\beta-1}$.

The same photons also prevent $\gamma$-rays from escaping the source. The $\gamma\gamma$ optical depth can be written in terms of $\varepsilon_\nu$ \citep[see Eq.~8 in][]{Murase:2015xka} as:
\begin{equation}
\tau_{\gamma\gamma}[\varepsilon_{\gamma\gamma-p\gamma}]\sim10^3 f_{p\gamma}[20\, \varepsilon_\nu],
\end{equation}
where $\varepsilon_{\gamma\gamma-p\gamma}\sim10~{\rm GeV}~(\varepsilon_\nu/300~{\rm TeV})$. 
The fact that $10-100$~GeV $\gamma$-rays were observed during the flare suggests that the neutrino production in the same emission region has to be inefficient \citep[e.g.,][]{Waxman:1997ti,Levinson:2006br,dermer_2007, Murase:2015xka,petro_2017}. 
Imposing $\tau_{\gamma\gamma}<1$ at 100~GeV leads to $f_{p\gamma}<{10}^{-3}~{(\varepsilon_p/60~{\rm PeV})}^{\beta-1}$. 
Note that the neutrino energy is related to the proton energy as $\varepsilon_\nu\approx0.05\varepsilon_p=0.3~{\rm PeV}~(\varepsilon_p/6~{\rm PeV})$.

For an isotropic-equivalent proton luminosity, $\varepsilon_{p}L_{\varepsilon_p}$, the differential neutrino luminosity is then given by
\begin{eqnarray}
\varepsilon_{\nu}L_{\varepsilon_\nu}&\approx&\frac{3}{8}f_{p \gamma}(\varepsilon_{p}L_{\varepsilon_p})\nonumber\\
&\simeq&1.2\times{10}^{45}~{\rm erg}~{\rm s}^{-1}~f_{p\gamma,-4}\left(\frac{\varepsilon_{p}L_{\varepsilon_p}}{{10}^{49.5}~{\rm erg}~{\rm s}^{-1}}\right),\,\,\,\,\,
\end{eqnarray} 
which is consistent with the results of \cite{Keivani:2018rnh}.  
The remaining fraction (i.e., 5/8) of energy should be carried by pionic $\gamma$-rays with $\varepsilon'_\gamma\approx0.1\varepsilon'_p$ and secondary electrons and positrons with $\gamma'_e\approx0.05\varepsilon'_p/(m_ec^2)\simeq2.9\times{10}^7~(\varepsilon_p/6~{\rm PeV})(20/\delta)$. The TeV-PeV $\gamma$-rays are attenuated inside the source, which also generate the pairs. The Bethe-Heitler process also injects high-energy pairs with $\gamma'_e\approx5\times{10}^{-4}\varepsilon'_p/(m_ec^2)\simeq2.9\times{10}^5~(\varepsilon_p/6~{\rm PeV})(20/\delta)$ \citep[e.g.][]{mastichiadis_1995}; even more energetic pairs can be produced by interactions happening far from the energy threshold of the process \citep[e.g.][]{kelner_2008}. These highly relativistic pairs quickly lose their energies via synchrotron and IC cooling. The cooling Lorentz factor is $\gamma_c'\approx2300~{B'}_{-0.5}^{-2}{l'_{17}}^{-1}{(1+Y_{\rm IC})}^{-1}$, implying that the resulting cascade spectrum lies in the fast-cooling regime. In the case of TXS 0506+056, the synchrotron peak is comparable to the IC peak, and the Compton Y parameter ($Y_{\rm IC}$) is at most unity~\citep{Keivani:2018rnh}.  

The synchrotron emission from pairs injected via the Bethe-Heitler process is not always negligible in blazars, as demonstrated by \cite{petro_2015}. It turns out to be important also for TXS 0506+056 during its high state \citep{Keivani:2018rnh}. 
The minimum synchrotron cascade flux associated with the neutrino flux at $\varepsilon_\nu$ is:  
\begin{eqnarray}
\varepsilon_{\gamma}L_{\varepsilon_\gamma}|_{\varepsilon_{\rm syn}^{\rm BH}}&\approx&\frac{1}{2(1+Y_{\rm IC})}g[\beta]f_{p\gamma}(\varepsilon_{p}L_{\varepsilon_p})\nonumber\\
&\approx&\frac{4g[\beta]}{3(1+Y_{\rm IC})}\varepsilon_{\nu}L_{\varepsilon_\nu},
\label{eq:BHsyn}
\end{eqnarray}
where $\varepsilon_{\rm syn}^{\rm BH}\simeq6~{\rm keV}~{B'}_{-0.5}{(\varepsilon_p/6~{\rm PeV})}^2(20/\delta)$ is the characteristic frequency of synchrotron emission by pairs from protons with $\varepsilon_p\approx20\varepsilon_\nu$. 
Because of the broad distribution of pairs injected by the Bethe-Heitler processes, even if the protons are monoenergetic, \citep{dimitrakoudis_2012} the expected synchrotron spectrum will be extending over several decades in energy~\citep[e.g.][]{petro_2015,Petropoulou:2015upa}.
Note that for sufficiently high-energy pairs we expect $Y_{\rm IC}\ll1$ due to the Klein-Nishina suppression.  

Similarly, for synchrotron emission from pairs injected via photomeson production and two-photon annihilation for pionic $\gamma$-rays, the synchrotron cascade flux is:
\begin{eqnarray}
\varepsilon_{\gamma}L_{\varepsilon_\gamma}|_{\varepsilon_{\rm syn}^{p\gamma}}&\approx&\frac{1}{2(1+Y_{\rm IC})}\frac{5}{8}f_{p\gamma}(\varepsilon_{p}L_{\varepsilon_p})\nonumber\\
&\approx&\frac{5}{6(1+Y_{\rm IC})}\varepsilon_{\nu}L_{\varepsilon_\nu},
\label{eq:pgsyn}
\end{eqnarray}
where $\varepsilon_{\rm syn}^{p\gamma}\simeq60~{\rm MeV}~{B'}_{-0.5}{(\varepsilon_p/6~{\rm PeV})}^2(20/\delta)$ and the contribution of pionic $\gamma$-rays is included assuming that they are converted into pairs inside the source. 


In addition to the synchrotron cascade components considered above, the IC emission and subsequent regeneration processes can affect the pair-injection spectrum. Although the exact spectral shape of a cascade photon spectrum depends on details of the pair injection and possible contributions from muon and meson radiation, the resulting energy spectrum  becomes approximately flat, that is, it can be expressed as $E_\gamma F_{E_\gamma}\propto E_\gamma^{2-\beta}$ with $\beta\sim1.5-2$ varying in the X-ray and $\gamma$-ray range. 

For the TXS 0506+056 flare coincident with IceCube-170922A, {\it Swift} and {\it NuSTAR} measured X-rays quasi-simultaneously. A more recent sophisticated analysis gave the X-ray flux, $E_\gamma F_{E_\gamma}^X\approx0.8\times{10}^{-12}~{\rm erg}~{\rm cm}^{-2}~{\rm s}^{-1}$ at $E_\gamma\approx2-3$~keV~\citep{Keivani:2018rnh}. 
This leads to tight limits on the high-energy neutrino flux from the TXS 0506+056 flare. 
Combining Eq.~(\ref{eq:BHsyn}) with the observed X-ray flux, the neutrino flux in the $0.1-1$~PeV range is constrained as $E_\nu F_{E_\nu}^{0.1-1~{\rm PeV}}\lesssim E_\gamma F_{E_\gamma}^X \sim{10}^{-12}$~\ergcmsqs~for all flavors, or we have $\varepsilon_\nu L_{\varepsilon_{\nu_\mu}}^{0.1-1~{\rm PeV}}\lesssim \varepsilon_\gamma L_{\varepsilon_\gamma}^X/3 \sim{10}^{44}~{\rm erg}~{\rm s}^{-1}$, where a factor of 3 comes from $\nu_e:\nu_\mu:\nu_\tau\approx1:1:1$. This is fully consistent with the detailed numerical results presented in \cite{Keivani:2018rnh}, and the upper limit neutrino luminosity is much lower than the luminosity suggested for the 2017 and 2014-2015 flares of TXS 0506+056, $\varepsilon_\nu L_{\varepsilon_{\nu}}(\approx3\varepsilon_\nu L_{\varepsilon_{\nu_\mu}})\sim{10}^{46}-{10}^{47}~{\rm erg}~{\rm s}^{-1}$. 
Note that the neutrino luminosity around the peak at higher energies~\citep[see Fig.~4 of][]{Keivani:2018rnh} is slightly higher owing to $g[\beta]$ and details of the cascade spectrum.  

Eqs.~(\ref{eq:BHsyn}) and (\ref{eq:pgsyn}) show that the luminosity of the synchrotron cascade components is comparable to the neutrino luminosity, as long as $Y_{\rm IC}\ll 1$. Thus, the cascade bound on the neutrino flux is unavoidable as long as the photomeson production occurs in a compact region such as the blazar zone.  
If the canonical picture of blazars based on the single-zone modeling is correct, these results allow us to predict that the 2017 and 2014-2015 neutrino flares reported by the IceCube Collaboration~\citep{Aartsen2018blazar2} should be accompanied by X-ray emission with $E_\gamma F_{E_\gamma}^X\sim E_\nu F_{E_\nu}^{0.1-1~{\rm PeV}}\sim(3-30)\times{10}^{-11}$~\ergcmsqs, which should be detectable by X-ray sky monitors such as {\it Swift} and {\it MAXI}.

Finally, we comment on high-energy neutrino emission through hadronuclear processes. 
For a relativistic jet of blazars, $pp$ interactions as the neutrino production mechanism cannot be globally efficient at least in the blazar zone; otherwise the jet becomes too heavy to remain relativistic. 
However, for flares, one could potentially avoid this problem by invoking some mechanism to entrain dense clouds in the jets.
Several possibilities have been discussed, which include clouds in the broad-line region, stellar winds, and the tidal stripping of the stellar envelope~\citep[e.g.,][]{Bednarek:1996ffa,BoschRamon:2012td,Perucho:2017iem}. In addition, massive stars could supply matter via supernova explosions or mass eruptions. 
Although various possibilities could potentially be considered (see Fig.~\ref{fig2}), there are several issues in the explanation for the month-scale neutrino flares from TXS 0506+056.
First, there has been much evidence that the clouds in the broad-line region are mostly located around the equatorial plane, not on the jet axis~\citep[e.g.,][]{1986ApJ...302...56W,2013MNRAS.429..135R,2014Natur.513..210S}. The profiles of the broad-line emission often show a disk-like rotation~\citep[e.g.,][]{2009NewAR..53..133E,2013MNRAS.429.1479L}, and the vertical velocity component can be naturally explained by the dusty disk wind~\citep[e.g.,][]{2011A&A...525L...8C,2015AdSpR..55.1806C}. On the other hand, there are a large number of stars around the AGN core, and some of them could be on the jet axis. 
No matter what the origin is, the cloud mass required to achieve a high column density of $\gtrsim{10}^{24}~{\rm cm}^{-2}$ is $M_c\gtrsim0.4~M_\odot~l_{c,16}^2$ for the cloud size $l_c$ (note that the observations suggest that the comoving size of the blazar zone is $l'\sim{10}^{16}-{10}^{17}$~cm). With a typical jet luminosity of BL Lacs, it is difficult for a massive cloud to become highly relativistic. If the CR generation occurs via collisions between the jet and a dense cloud at rest, the characteristic time scale is given by $(1+z)l_c/\delta^2c\simeq1.1\times{10}^4~{\rm s}~l_{c,17}{(20/\delta)}^2$, which is shorter than the observed variability and duration, although the evaporation and expansion subsequent to the jet-cloud interaction may eventually lead to a longer time scale.
Third, the existence of a powerful jet implies that such a cloud can be largely ionized and strong photoelectric absorption of X-rays is unlikely to happen except for the innermost region of the cloud. 
For a cloud density of $n_c\simeq2.8\times{10}^6~{\rm cm}^{-3}~(M_c/10~M_\odot)l_{c,17}^{-3}$, the ionization radius is estimated to be $\sim8\times{10}^{17}~{\rm cm}~{(\varepsilon_\gamma L_{\varepsilon_\gamma}/10^{45}~{\rm erg}~{\rm s}^{-1})}^{1/3}{(\varepsilon_\gamma/100~{\rm eV})}^{-1/3}n_{c,7}^{-2/3}{\mathcal T}_{5}^{1/4}$, where $\mathcal T$ is the cloud temperature. Except for extreme cases with $N_H\gg\sigma_T^{-1}$, we expect that X-rays escape and the cascade limit is important even for $pp$ models, and we have: 
\begin{eqnarray}
\varepsilon_{\gamma}L_{\varepsilon_\gamma}|_{\varepsilon_{\rm syn}^{pp}}&\approx&\frac{1}{2(1+Y_{\rm IC})}\frac{1}{2}f_{pp}(\varepsilon_{p}L_{\varepsilon_p})\nonumber\\
&\approx&\frac{1}{2(1+Y_{\rm IC})}\varepsilon_{\nu}L_{\varepsilon_\nu}.
\label{eq:ppsyn}
\end{eqnarray}
For a spectrum of CRs with $s_{\rm cr}\sim2$, the above relation holds at energies from $\varepsilon_{\rm syn}^{c}\simeq0.39~{\rm eV}~{B'}_{-0.5}^{-3}{l'}_{17}^{-2} {(1+Y_{\rm IC})}^{-2}(\delta/20)$ to $\varepsilon_{\rm syn}^{pp,\rm max}\simeq60~{\rm MeV}~{B'}_{-0.5}{(\varepsilon_p^{\rm max}/6~{\rm PeV})}^2(20/\delta)$.

\subsection{Multizone Models and CR-induced Neutral Beams}
An electromagnetic cascade is a consequence of energy conservation, so the cascade limit discussed in the previous subsection exists for not only single-zone but also multizone models. 

One of the possibilities is that neutrinos are mainly produced around the base of the jet or even in the vicinity of a supermassive black hole (i.e., AGN core models)~\citep[e.g.,][]{Stecker:1991vm,bednarek_1999,becker_2009,Stecker:2013fxa,Kimura:2014jba}. 
This includes high-energy neutrino production caused by $pp$ interactions with dense clouds as discussed above. 
Because of the higher compactness of the system, broadband cascade emission should accompany the neutrino emission. In such an inner region, magnetic fields should also be much stronger, which could affect the resulting cascade spectrum. Importantly, not only protons but also electrons are accelerated, and both emission components should be taken into account. 
Detailed studies are beyond the scope of this work (Mastichiadis \& Petropoulou 2018 in preparation).

\begin{figure}[t]
\includegraphics[width=\linewidth]{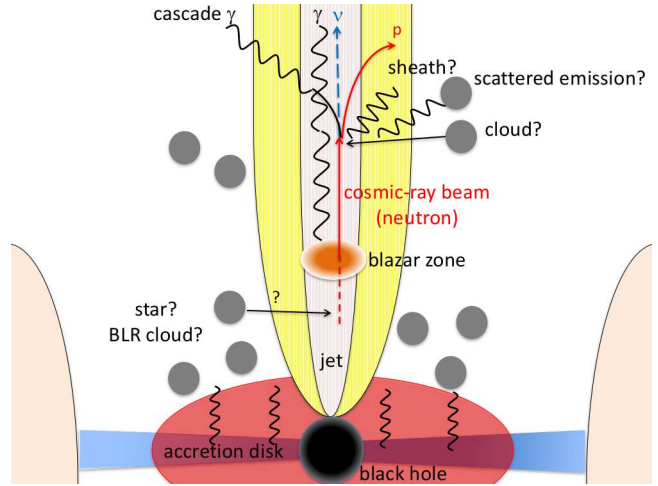}
\caption{Schematic picture (not in scale) of the CR-induced beam model for high-energy neutrino production. See text for details~\citep[see also][]{Murase:2014foa,Dermer:2012rg}. 
While the neutrino emission is highly beamed, the associated cascade emission in the X-ray range is isotropized.
\label{fig2}
}
\end{figure}

Here we consider the CR-induced neutral beam model (see Fig.~\ref{fig2}), which can avoid the cascade constraints.
We consider interactions between beamed CRs escaping from the blazar zone and an external radiation field. Although we do not specify the origin of the external photons, this setup is analogous to the one considered in 
\cite{Murase:2014foa} and \cite{Dermer:2014vaa}. In this sense, this two-zone model can be regarded as a natural extension of the standard leptonic scenario. 
In particular, we assume that escaping CRs are neutrons that can be produced via the photodisintegration of nuclei in the blazar zone. 
The $\gamma$-ray signatures produced by CR-induced neutral beams were previously studied in \cite{Murase:2011yw} and \cite{Dermer:2012rg} \citep[see also][for related discussion about intergalactic cascades]{Essey:2009ju,Essey:2010er,Murase:2011cy}. 
As we show below, the cascade emission can be largely diminished via the isotropization of relativistic electrons and positrons.  
 
Following \cite{Dermer:2014vaa}, let us assume that CRs are accelerated via the second-order Fermi acceleration mechanism. The maximum energy accelerated in the blazar zone can be $\varepsilon_{\rm cr}/Z\sim1-10$~PeV, where $\varepsilon_{\rm cr}$ is the CR ion energy and $Z$ is the nuclear charge. The CR acceleration zone can be the $\gamma$-ray emission site or inner regions of the blazar zone, and disintegrated nuclei are accompanied by not only protons but also neutrons~\citep[e.g.,][]{Murase:2010va,Rodrigues:2017fmu}. 
The protons may lose energy via adiabatic losses during the confinement in the blazar zone, while neutrons can escape. 
The neutron luminosity is given by $\varepsilon_nL_{\varepsilon_n}\approx(1/2)f_{A\gamma}(\varepsilon_{\rm cr}L_{\varepsilon_{\rm cr}})$, where $f_{A\gamma}$ is the effective optical depth to the photodisintegration process~\citep{Murase:2010va} and $\varepsilon_{\rm cr}L_{\varepsilon_{\rm cr}}$ is the CR ion luminosity.  
For neutron production in the $\gamma$-ray emission region, we have $f_{A\gamma}\sim0.1~{(\varepsilon_\gamma L_{\varepsilon_{\gamma}}/10^{46}~{\rm erg}~{\rm s}^{-1})}{(\delta/20)}^{-3}{l'}_{17}^{-1}{(\varepsilon_\gamma/1~\rm eV)}^{-1}$\\${(\varepsilon'_{\rm cr}/\tilde{\varepsilon}_{A\gamma,\rm syn}^{\prime})}^{\beta-1}$, where $\tilde{\varepsilon}_{A\gamma,\rm syn}^\prime=0.5m_Ac^2\bar{\varepsilon}_{\rm GDR}/\varepsilon'_{\rm syn}$ and $\bar{\varepsilon}_{\rm GDR}\sim20-30$~MeV. This also implies that the neutron emission may predominantly come from smaller dissipation radii at which efficient photodisintegration (i.e., $f_{A\gamma}\gtrsim1$) occurs.

In single-zone models, the neutrino flares of TXS 0506+056 require unpleasantly large CR luminosities~\citep{Keivani:2018rnh}. This problem still exists at some level even in multizone models, although it can be alleviated in the CR beam model in the sense that the meson production efficiency is enhanced by additional target photons or nucleons.   
One should keep in mind that observations and modeling of radio galaxies (based on larger-scale jets than the blazar zone) have shown that the absolute jet power averaged over the lifetime of the AGN jet is $P_j\lesssim{10}^{45}-{10}^{46}~{\rm erg}~{\rm s}^{-1}$ for Fanaroff-Riley I galaxies~\citep{2010ApJ...720.1066C,Godfrey:2013kt}, which are are believed to be off-axis counterparts of BL Lacs.  
For the supermassive black hole mass $M_{\rm BH}$, the Eddington luminosity\footnote{The X-ray observations~\citep{Keivani:2018rnh} indicate that the disk luminosity in the X-ray range has to be lower than $3\times{10}^{44}~{\rm erg}~{\rm s}^{-1}$, which is consistent with the common belief that BL Lacs are associated with radiatively inefficient accretion disks.} is $L_{\rm Edd}\simeq1.3\times{10}^{47}~{\rm erg}~{\rm s}^{-1}~{(M_{\rm BH}/10^9~M_\odot)}$. \cite{Ghisellini:2014pwa} showed that the absolute jet power of blazars may exceed the accretion luminosity, and our study implies that the flaring jet power is larger than the time-averaged one by $b_{\rm fl}/f_{\rm fl}\sim3-10$. The isotropic-equivalent CR luminosity during the flaring phase can then be written as $L_{\rm cr}^{\rm fl}\approx(2/\theta_{\rm beam}^2)\epsilon_{\rm cr}P_{j}(b_{\rm fl}/f_{\rm fl})\simeq6.0\times{10}^{49}~{\rm erg}~{\rm s}^{-1}~{(\theta_{\rm beam}/0.05)}^{-2}(\epsilon_{\rm cr}/0.2)(b_{\rm fl}f_{\rm fl}^{-1}/10)$\\$(P_j/0.3L_{\rm Edd})(M_{\rm BH}/{10}^9~M_\odot)$, where $\epsilon_{\rm cr}$ is the energy fraction carried by CR ions and $\theta_{\rm bream}$ is the opening angle of the CR beam. The neutron luminosity during the flaring phase results in $L_n^{\rm fl}\simeq3.0\times{10}^{49}~{\rm erg}~{\rm s}^{-1}~f_{A\gamma}{(\theta_{\rm beam}/0.05)}^{-2}(\epsilon_{\rm cr}/0.2)(b_{\rm fl}f_{\rm fl}^{-1}/10)$\\$(P_j/0.3L_{\rm Edd}){(M_{\rm BH}/10^9~M_\odot)}$.

The neutrons that leave the CR acceleration zone propagate along the jet and may interact with external radiation fields that could exist on larger scales or perhaps a dense cloud.
For LSP and ISP objects like TXS 0506+056, it is possible to invoke such a setup. 
For example, if the jet is structured, nonthermal photons can be provided by the sheath region. Moreover, a fraction of UV and X-ray emission from the accretion disk can be scattered by clumps of matter that may be present at outer radii. In addition, there could be high-velocity clumps such as the broad-line region, although they are usually seen in FSRQs. 
Note that the neutrons with $\gamma_n\sim{10}^7-{10}^8$ can travel $\sim0.1-1$~kpc.  

Interestingly, the detailed modeling of the SED of TXS 0506+056~\citep{Keivani:2018rnh} already suggested that such an external radiation field is necessary to explain the X-ray and $\gamma$-ray spectrum. If this is the case, it is natural for escaping CRs to keep interacting with the ambient photons, leading to the production of more neutrinos. 

As a toy model, we assume that the decelerated jet or slower jet of the sheath region provides soft photons with a luminosity of $L_{\rm ext}\sim3\times{10}^{45}~{\rm erg}~{\rm s}^{-1}$ and the characteristic energy at $\varepsilon_{\rm ext}\sim10$~eV, over a length scale of $R_{\rm ext}\sim3\times{10}^{19}$~cm.  
The external radiation energy density is $U_{\rm ext}\approx 3L_{\rm ext}/(4\pi R_{\rm ext}^2c)\sim3\times{10}^{-5}~{\rm erg}~{\rm cm}^{-3}$, which is consistent with the parameters used in \cite{Keivani:2018rnh}. 
Noting $\hat{\sigma}_{n\gamma}\approx\hat{\sigma}_{p\gamma}$, the photomeson production efficiency is $f_{n \gamma}\approx[\eta_{n\gamma}\hat{\sigma}_{n\gamma}3L_{\rm ext}/(4\pi R_{\rm ext}c\varepsilon_{\rm ext})]
{(\varepsilon_n/\tilde{\varepsilon}_{n\gamma,\rm ext})}^{\beta-1}
\sim3\times{10}^{-3}~\eta_{n\gamma}L_{\rm ext,45.5}R_{\rm ext,19.5}^{-1}{(\varepsilon_{\rm ext}/10~\rm eV)}^{-1}{(\varepsilon_n/\tilde{\varepsilon}_{n\gamma,\rm ext})}^{\beta-1}$ (where $\eta_{n\gamma}=\eta_{p\gamma}$ and $\tilde{\varepsilon}_{n\gamma,\rm ext}=0.5m_nc^2\bar{\varepsilon}_{\Delta}/\varepsilon_{\rm ext}\approx\tilde{\varepsilon}_{p\gamma,\rm ext}$),
which can be larger by a factor of $\sim R_{\rm ext}/(\Gamma l')$ compared to $f_{p\gamma}$ in the blazar zone (see Eq.~\ref{eq:fpgamma}). 
Alternatively, instead of $n\gamma$ interactions, one could expect $np$ interactions between the neutral beam and a dense cloud in the line of sight. Although the existence of such a cloud is speculative, with a clump like a giant molecular cloud with a mass of $M_c\sim({10}^{5}-{10}^6)~M_\odot$ and a size of $l_c\sim{10}^{19}$~cm, the effective $np$ optical depth can be as high as $f_{np}\simeq0.011~(M_{\rm c}/10^5~M_\odot)l_{c,19}^{-2}$.  

The all-flavor neutrino luminosity is:
\begin{eqnarray}
\varepsilon_{\nu}L_{\varepsilon_{\nu}}\simeq3.8\times{10}^{46}~{\rm erg}~{\rm s}^{-1}~\left(\frac{2K}{1+K}\right)\left(\frac{f_{n\gamma/np}\varepsilon_{n}L_{\varepsilon_n}}{{10}^{47}~{\rm erg}~{\rm s}^{-1}}\right),\,\,\,\,\,\,\,\,\,\,\,
\end{eqnarray}
where $K=1$ and $K=2$ for $np$ and $n\gamma$ interactions, respectively.  
In the $n\gamma$ case, the neutrino energy corresponding to the neutron energy with $\tilde{\varepsilon}_{n\gamma,\rm ext}$ is:
\begin{equation}
\varepsilon_\nu\approx0.05\tilde{\varepsilon}_{n\gamma,\rm ext}\sim0.7~{\rm PeV}~{(\varepsilon_{\rm ext}/10~{\rm eV})}^{-1},
\end{equation}
which could explain both of the 2017 and 2014-2015 neutrino flares of TXS 0506+056, e.g., with $f_{A\gamma}\sim0.1$ and $f_{A\gamma}\gtrsim1$, respectively. 
The duration of neutrino emission is comparable to the ``lifetime'' of the CR-induced beam, which is being determined by the duration of particle energization in the CR acceleration zone corresponding to the observed $t_{\rm dur}$. 

The key point of the CR-induced neutral beam model considered here is that the cascade signature except for $\gamma$-rays in the very high-energy range can be largely diminished because of the isotropization of the relativistic pairs in the larger-scale jet or other magnetized environments. 
Let us assume that the magnetic field in the main scale of neutrino production is as small as $B_{\rm ext}\sim0.1-10$~mG (typical for large-scale jets). The deflection of pairs occurs before they cool via synchrotron and IC losses. 
Following \cite{Murase:2011yw}, the deflection angle during the radiation cooling time is given by:
\begin{equation}
\theta_{\rm def}\approx\sqrt{\frac{2}{3}}\frac{ct_{\rm syn}}{r_L}
\simeq3.5~\gamma_{e,9}^{-2}B_{\rm ext,-2.5}^{-1},
\end{equation}
where $t_{\rm syn}$ is the synchrotron cooling time, $r_L$ is the Larmor radius, and the Lorentz factor of pairs in the black hole rest frame is $\gamma_e\approx0.05\varepsilon_n/(m_ec^2)\simeq5.8\times{10}^8~(\varepsilon_n/6~{\rm PeV})$. 
Only synchrotron cooling is considered, because the IC cooling is suppressed for such high-energy electrons and positrons. The above equation implies that the pairs lose their energy after they become isotropized.  
Note that the X-ray emission is doubly suppressed, because the deflection is larger than the jet opening angle, i.e., $\theta_{\rm def}\gg\theta_j$, and the time spread in the cascade emission is longer than the intrinsic flare duration, i.e., $R_{\rm ext}/c\gg t_{\rm dur}$. 
Also, the CR neutrons do not produce extra pairs via the Bethe-Heitler process. In addition, in the $np$ scenario, X-rays could also be isotropized via the Compton scattering. Thus, X-rays from the neutral beam can be highly suppressed if the CR energy is lower than UHE energies, and the associated hadronic cascade signatures can be masked by the GeV $\gamma$-ray emission from the blazar zone. 
Note that the UHE neutron beam, if it exists, can also produce beamed cascade emission at $\gtrsim0.1$~TeV energies. The resulting synchrotron pair-echo emission~\citep{Murase:2011yw,Dermer:2012rg} is expected to have some time delay. We also remark that the very high-energy flare seen by MAGIC detected about 10 days later after IceCube-170922A.

\section{Summary}\label{sec:summary}
We considered implications of the high-energy neutrino flares from TXS 0506+056 by examining various constraints. 
Flaring blazars could be the brightest sources in the neutrino sky, while the observations can most naturally be reconciled with existing theoretical models, if the blazars are subdominant ($\sim1-10$\%) in the diffuse neutrino intensity. 
Interestingly, within the standard leptonic scenario of $\gamma$-ray emission, we found that the blazar neutrino emission itself can readily be dominated by flaring episodes. 
This could explain why the significant neutrino signals from TXS 0506+056 were found only in the time-dependent search and that the steady component that typically dominates $\gamma$-ray emission has not been seen yet in neutrinos.  
Bright neutrino flares like the ones observed for TXS 0506+056 could contribute only up to a few percent of the INB. If the association with this blazar is physical, such flares can be detected with a rate of $\lesssim1~{\rm yr}^{-1}$ by dedicated time-dependent searches in the near future. 

Based on analytical considerations, we also showed the importance of X-ray constraints to test the physical models of TXS 0506+056. An efficient electromagnetic cascade is unavoidable in the canonical blazar models based on the single radiation zone, which also predicts that the 2014-2015 neutrino flare found in the archival data should be accompanied by X-ray emission with $E_\gamma F_{E_\gamma}^X\sim(3-30)\times{10}^{-11}$~\ergcmsqs. This can readily be tested by X-ray sky monitors such as {\it Swift} and {\it MAXI}. Nondetection of the X-ray flares will necessitate more complicated models involving multizone emission. 
As a possible example, we discussed the CR-induced neutral beam model. In this model, neutrino production is expected to occur via, e.g., photomeson production on external radiation fields, which were also inferred by the detailed modeling of the SED of TXS 0506+056~\citep{Keivani:2018rnh}. Remarkably, the cascade emission, which is unavoidable whether neutrinos are produced by either photohadronic or hadronuclear process, can be largely diminished by the isotropization in magnetized environments and the absence of the Bethe-Heitler process, so that the X-ray constraints can be satisfied. 


The reported significance of the neutrino flare from TXS 0506+056, $3\sigma-4\sigma$, is intriguing. However, the observed coincidence still lacks convincing explanations in view of the multi-messenger data. More observational and theoretical efforts are necessary to confirm whether flaring blazars are the sources of high-energy neutrinos.  

Even if blazars are established as the sources of neutrinos by future observations, our results imply that it will not address the two most important questions, that is, which source class the main origin of high-energy cosmic neutrinos is, and where UHECRs come from. Other sources or populations (or perhaps different regions) are most likely to be responsible for the bulk of high-energy cosmic neutrinos and UHECRs. Thus, next generation neutrino detectors such as IceCube-Gen2~\citep{Aartsen:2014njl} and KM3Net~\citep{Adrian-Martinez:2016fdl} will be necessary to address these puzzles. 


\begin{acknowledgements}
We thank Derek Fox, Kunihito Ioka, Albrecht Karle, Nobuyuki Kawai, Azadeh Keivani, Shigeo Kimura, Peter M\'esz\'aros, Masaru Shibata, Yutaro Tachibana, Michael Unger, and Shigeru Yoshida for useful discussions. K.M. also acknowledges Niel Brandt and Amir Levinson for the discussions about the geometry of broad-line regions and jet energetics, respectively. The work of K.M. is supported by the Alfred P. Sloan Foundation and NSF grant No. PHY-1620777. F.O. acknowledges support from the DFG through grant SFB1258 ``Neutrinos and Dark Matter in Astro- and Particle Physics". M.P. acknowledges support from the Lyman Jr.~Spitzer Postdoctoral Fellowship.
This research was supported by the Munich Institute for Astro- and Particle Physics (MIAPP) of the DFG cluster of excellence ``Origin and Structure of the Universe''.
\end{acknowledgements}

\bibliographystyle{aasjournal}
\bibliography{kmurase}

\end{document}